\documentstyle[prb,aps,multicol,epsfig]{revtex}

\begin{document}
\draft
\title{Shadow features and shadow bands in the paramagnetic state\\
of cuprate superconductors}
\author{Y. M. Vilk}
\address{Materials Science Division, Argonne National Laboratory,\\
Argonne, IL 60439\\
           e-mail: yvilk@hexi.msd.anl.gov}
\maketitle

\begin{abstract}
\leftskip 54.8pt \rightskip 54.8pt The conditions for the precursors of
antiferromagnetic bands in cuprate superconductors are studied using
weak-to-intermediate coupling approach. It is shown that there are, in fact,
three different precursor effects due to the proximity to antiferromagnetic
instability: i) the shadow band which associated with new pole in the
Green's function ii) the dispersive shadow feature due to the thermal
enhancement of the scattering rate and iii) the non-dispersive shadow
feature due to quantum spin fluctuation that exist only in $\vec{k}-$scan of
the spectral function $A(\omega _{Fixed},\vec{k})$. I found that dispersive
shadow peaks in $A(\omega,\vec{k})$ can exist at {\em finite} temperature $T$
in the renormalized classical regime, when $T\gg \omega _{sf}$, $\xi_{AFM}
>\xi_{th}=v_F/T$ ($\omega _{sf}$ is the characteristic energy of spin
fluctuations, $\xi_{th}$ is the thermal wave length of electron). In
contrast at zero temperature, only non-dispersive shadow feature in $%
A(\omega_{Fixed},\vec{k})$ has been found. I found, however, that the latter
effect is always very small. The theory predict no shadow effects in the
optimally doped materials. The conditions for which shadow peaks can be
observed in photoemission are discussed.
\end{abstract}


\begin{multicols}{2}   
\narrowtext  

\section{Introduction}

Recent advances in the Angular Resolved Photoemission technique (ARPES) have
allowed to determine electronic excitations of some High-$T_c$ materials.
Two different approaches to photoemission are currently used. In type I
ARPES one measures $A(\omega ,\vec{k}_{Fixed})$ the probability that an
excitation of fixed momentum $\vec{k}_{Fixed}$ has an energy $\omega .$ In
the type II ARPES one measures the probability $A(\omega _{Fixed},\vec{k})$
that an excitation of fixed energy $\omega _{Fixed}$ has a momentum $\vec{k}$%
{\bf . }Using the latter technique on the high-temperature superconductor
B2212, Aebi et al.\cite{Aebi} have found, in addition to the main Fermi
surface (F. S.), another F. S. that is identical to the first one but of
much smaller intensity and displaced from the main F. S. by the
antiferromagnetic (AFM) wave vector $\vec{Q}=(\pi /a,\pi /a)$ . This
structure has been taken as evidence for the existence of the ``shadow
bands'' that had been predicted much earlier by Kampf and Schrieffer \cite
{KShr}. These authors had argued that precursors of the AFM bands should
appear in the paramagnetic state close to the AFM instability. More
recently, observation of dispersing shadow bands has been also reported in
type I photoemission experiments \cite{Rosa}. However, the AFM
interpretation of the phenomenon observed in B2212 remains controversial
because: i) the AFM correlation length $\xi $ in optimally doped B2212 is
rather short \cite{Chakravarty}, ii) the effect has not been found yet in
any other high-temperature superconductor or in the underdoped B2212 iii) it
becomes more pronounced when one increase orthorombicity by lead doping \cite
{Aebi2}. All these leave open the possibility of structural origin of the
phenomenon observed in Refs. \cite{Aebi,Rosa}. It is thus very important to
understand under which general conditions shadow bands can occur in
practice. Some aspects of this issue have been considered in a number of
recent theoretical publications\cite{VT1}-\cite{Haas}. In particular, the
strong coupling numerical calculations for the t-J model \cite{Haas} show
peaks outside non-interacting F. S. in the underdoped case for $T=0$.
However, these results have been obtained so far only for very small
clusters and the results close to optimal doping are unclear \cite{Haas}.

In this paper, I study under which general conditions shadow bands can
occur, using a well known method that can be justified up to intermediate
coupling. I show that there are, in fact, a few different precursors effects
due to the proximity to AFM instability depending on whether the effect is
due to classical $\omega _{sf}\ll T$ or quantum $T\ll \omega _{sf}$ spin
fluctuations (here $\omega _{sf}$ is characteristic energy of spin
fluctuations). I also show that shadow effects in type I and type II
photoemission experiments may occur under different conditions. For type I
ARPES experiments, I find that while classical spin fluctuations can produce
dispersive shadow effects in $A(\omega ,\vec{k})$, quantum spin fluctuations
cannot. The key difference between both regimes is that classical thermal
fluctuations lead to precursors of the magnetic Bragg peaks in the {\em %
static} spin structure factor, while the dynamical nature of quantum spin
fluctuations does not allow sharp structures in $S_{sp}(\vec{q})$, no matter
how large $\xi /a$ is ($a$ lattice constant). Consequently, shadow effects
will appear in type I experiments at low temperature only if the AFM
correlation length $\xi $ has the appropriate temperature dependence, not
just when $\xi /a$ is large enough. This conclusion is qualitatively
different from the results of numerical studies \cite{Deisz}-\cite{Haas}, in
which it was tacitly assumed that the effect is determined by the parameter $%
\xi /a$. alone. It is only in type II ARPES experiments that quantum spin
fluctuations can lead to a non-dispersive shadow feature in $A(\omega
_{Fixed},\vec{k})$\cite{Chubukov}. I find, however, that the latter effect
is very small and exists only in a narrow range near zero energy. More
experimental implications of my results are discussed in the conclusion.

\section{Model and calculational procedure}

I evaluate the effect of spin fluctuations on the electronic self-energy in
the one-loop approximation. The expression for the self-energy has the form: 
\begin{equation}  \label{SigReal}
\begin{array}{c}
\Sigma (\omega , \vec k)=g \bar{g}\int \frac{d^2qd\omega ^{\prime }}{(2\pi
)^3} \\ 
\times \frac{\chi _{sp}^{\prime \prime }(\omega ^{\prime },\vec q%
)\left[\coth \left( \frac{\omega ^{\prime }}{2T}\right) -\tanh \left( \frac{%
\tilde \varepsilon ( \vec k+\vec q)}{2T}\right) \right]}{\omega -\omega
^{\prime }-\tilde \varepsilon ( \vec k+\vec q)+i0}
\end{array}
\end{equation}
where $\chi _{sp}(\omega ,\vec q)$ is the spin susceptibility, $g\bar{g}$ is
the effective coupling constant between electrons and spin fluctuations and $%
\tilde \varepsilon (\vec k)$ is the energy dispersion relative to the
chemical potential. The constant vertex corrections are included in $\bar{g}$%
. It was shown within the Hubbard model\cite{VT1}, that by including vertex
corrections in this way the above self-energy expression (\ref{SigReal})
remains a good approximation up to intermediate coupling.

Close to AFM instability the low-energy asymptotic form of the spin
susceptibility $\chi _{sp}$ can be written in the Orstein-Zhernike form: 
\begin{equation}
\chi _{sp}(\omega ,\vec{q})=\frac{\chi _{sp}(0,\vec{Q})}{1+(\vec{q}-\vec{Q}%
)^2\xi ^2-i\omega /\omega _{sf}}  \label{Susc1}
\end{equation}
where $\xi $ is the AFM correlation length and $\omega _{sf}$ is the
characteristic energy of the spin fluctuations. Any RPA-like theory, whether
it uses bare or renormalized vertices \cite{VT1} or propagators \cite{FLEX},
have the Orstein-Zhernike form close to $\vec{q}=\vec{Q}$. In the context of
High-$T_c$ materials, the Orstein-Zhernike susceptibility with particular
set of parameters is often called MMP susceptibility. It has been used to
fit a number of experiments, that are sensitive to the strong enhancement of 
$\chi $ around $\vec{q}=\vec{Q}$ \cite{MMP}. Here, I will adopt a more
general approach and consider for which general conditions on the
parameters $\xi $ and $\omega _{sf}$ the precursors of AFM bands can exist.
Since shadow effects come from the peak in $\chi $ around $\vec{q}=\vec{Q}$,
the Orstein-Zhernike form of $\chi $ is sufficient for my purposes and its
simplicity will allow me to obtain a number of results analytically. The
imaginary part of the susceptibility $\chi _{{sp}}^{\prime \prime }(\omega ,%
\vec{q})$ has to be cut off at some high frequency $\Gamma _{cut}$ in order
to satisfy the local moment sum rule. In electronic models this cutoff is of
the order of a few hopping integrals $t.$ For the low energy physics the
precise value of $\Gamma _{cut}$ is not important. I use $\Gamma _{cut}=5.5t$%
.

From a microscopic point of view (see, for example,\cite{MMP}), the
parameters in the susceptibility Eq.(\ref{Susc1}) scale with the correlation
length as: $\chi _{sp}(\vec{Q},0)\propto \xi ^2$, $\omega _{sf}\propto 1/\xi
^2$. Consequently the model is defined by only three parameters $\xi $, $%
\omega _{sf}\,$ and $g^{\prime }\equiv g\bar{g}\chi _{sp}(\vec{Q})/\xi ^2$.
Qualitatively, our results mainly depend on $\xi $ and on $T/\omega _{sf}$
while the parameter $g^{\prime }$ (in the Hubbard model \cite{VT1} $%
g^{\prime }\propto U$) determines the quantitative scale of the effects. To
model the band-structure of the hole-doped high-Tc materials I use the
simplest tight-binding model that correctly reproduces the experimentally
observed Fermi surfaces in YBCO and B2212 at optimal doping, namely $%
\epsilon (\vec{k})=-2t(\cos (k_x)+\cos (k_y))-4t^{\prime }\cos (k_x)\cos
(k_y)$ with $t^{\prime }=-0.45t$ (I set $a=1$) . The band structure estimate
for $t$ is about $t\approx 250meV$. Using this approach, I first present
numerical results and then analytically find the conditions for the
existence of the various effects.

Both the self-energy and the spectral function $A(\omega ,\vec k)\equiv
-2\Sigma ^{\prime \prime }(\omega ,\vec k)/[(\omega -\epsilon (\vec k)+\mu
-\Sigma ^{\prime }(\omega ,\vec k))^2+|\Sigma ^{\prime \prime }(\omega ,\vec %
k)|^2]$ are discussed. The chemical potential of the interacting electrons
is found from the usual condition $\int f(\omega )A(\omega ,\vec{k})d\omega
d^2k/(2\pi )^3=n/2$. I choose $n=0.83$ that corresponds to optimal doping in
B2212 \cite{Note1}.

\section{Results}

I start with numerical results  and figures that illustrate  various
points I wish to make. Analytical results for  various regimes
identified in the numerical examples appear in a subsequent section.

\subsection{Illustrative examples}

The important aspect of critical AFM fluctuations is that they lead to
strongly $\vec k$-dependent self-energy. The maximum of the scattering rate $%
-2\Sigma ^{\prime \prime }(0,\vec k)$ at finite temperature occurs on the
so-called shadow Fermi surface (Sh. F. S.). The latter is {\it defined} here
as the set of points obtained by the translation of the real F. S. by the
wave vector $\vec Q=(\pi,\pi)$. The inset in Fig.\ref{cross}(a) shows the
real F.S. and the Sh. F. S. for the optimally doped materials. The points at
which the real F. S. and the Sh. F. S. intersect are the points at which the
spectral function $A(\omega ,\vec k_F)$ is affected the most by spin
fluctuations. I start the analysis with these crossing points, $\vec k{\bf =}%
\vec k_{cr},$ also known as hot spots.

\subsubsection{{ Shadow bands at the hot spots}}

The filled line on Fig.\ref{cross} shows the spectral function $A(\omega ,%
\vec{k}_{cr})$ in the regime dominated by classical thermal fluctuations ($%
\omega _{sf}\ll T$ , $T=0.03t$ and $\xi =20a$). We see that instead of the
usual quasi-particle peak at $\omega =0$ there are two precursors of the
antiferromagnetic bands with a pseudogap between them. These bands are
associated with two new quasiparticle solutions of the equation $\omega
-\epsilon (\vec{k})+\mu -\Sigma ^{\prime }(\omega ,\vec{k})=0$. It is
because these solutions exist that the term {\it bands} is used. I also find
that in this case the scattering rate $\Sigma ^{\prime \prime }(\omega ,\vec{%
k}_{cr})$ has maximum at the Fermi level $\omega =0$.

I will show now by contradiction, that this effect exists only if $\xi (T)$
increases sufficiently rapidly with decreasing temperature. Indeed, suppose
the correlation length does not increase when temperature decreases below
the temperature $T=0.03$ used for the filled line on Fig.\ref{cross}. Taking
a much smaller temperature $T/t=0.005$ while keeping all other parameters
fixed, the dashed line on Fig.\ref{cross} shows that the pseudogap in $%
A(\omega ,\vec{k}_{cr})$ disappears while the quasiparticle peak reappears.
I find similar results when I keep the temperature and $\xi $ constant but
increase $\omega _{sf}$ up to $\omega _{sf}\approx T$ . This means that the
pseudogap in $A(\omega ,\vec{k}_{cr})$ can exist only if the correlation
length $\xi (T)$ increases sufficiently rapidly at low temperature. I show
below that $\xi (T)\propto \exp \left( const/T\right) $ is a sufficient
condition. The latter condition is realized in the renormalized classical
regime (R. C.) $\omega _{sf}\propto \xi ^{-2}\ll T$ that always precedes
the $T=0$ phase transition in two dimensions.

\subsubsection{{Shadow features in the diagonal direction}$(0,0)-(\pi ,\pi )$}

In this direction, real and shadow F.S. are well separated and $|\epsilon (%
\vec{k})-\epsilon (\vec{k}+\vec{Q})|$ is large. The filled curve in Fig.\ref
{Diag1} shows the spectral function at $T=0.03t$ for a wave vector $\vec{k}%
=(0.66,0.66)\pi $ that is just a little bit inside the Sh.~F.~S. ($\vec{k}%
_{Sh.F}\simeq (0.63,0.63)\pi $). The quasiparticle peak in this case is at
large positive energies and cannot be seen in the photoemission experiments
since they measure $f(\omega )A(\omega ,\vec{k})$. There is no new
quasiparticle solution to $\omega -\epsilon (\vec{k})+\mu -\Sigma ^{\prime
}(\omega ,\vec{k})=0$ in this case. Nevertheless, there is a local maximum
at negative binding energies (shown by arrow) that coincides with the
maximum in the imaginary part, $\Sigma ^{\prime \prime }(\omega ,\vec{k})$
in Fig.~\ref{Diag1}b. I call this a shadow {\it feature. }This maximum
occurs because for large values of $\left| \omega -\varepsilon (\vec{k})+\mu
\right| ,$ the quantity $\omega -\epsilon (\vec{k})+\mu -\Sigma ^{\prime
}(\omega ,\vec{k})$ is large and slowly varying in the region where $\Sigma
^{\prime \prime }(\omega ,\vec{k})$ acquires a maximum. Hence, this maximum
is directly reflected in the spectral function that becomes proportional to $%
\Sigma ^{\prime \prime }(\omega ,\vec{k})$ rather than inversely
proportional to it as in the case of quasiparticles. This maximum occurs at $%
\omega \approx \tilde{\varepsilon}(\vec{k}+\vec{Q})$, a dispersion that
looks like the AFM band but with the gap equal to zero. Nevertheless, this
is definitely {\it not} a real AFM band since a) it is thermally excited
(see below) b) it exists only in certain regions of $\vec{k}$-space where $%
|\epsilon (\vec{k})-\epsilon (\vec{k}+\vec{Q})|\gg \Sigma ^{\prime \prime
}(\omega ,\vec{k})$. Note that similar shadow features have been found in
the Hubbard model with $t^{\prime }=0$, using the FLEX approximation\cite
{FLEX}.

To show that this shadow feature is a thermally excited state I plot the
dashed line in Fig.\ref{Diag1} for the case where the temperature is
decreased to $T=0.003t$ with all other parameters the same as for $T=0.03t$.
Clearly the shadow feature (shown by arrow) disappears at lower temperature.
It also disappears for $\omega _{sf}>T$. I note that the maximum in $%
A(\omega ,\vec{k})$ around $\omega _{max}\sim -t$ is different from the
``shadow band'' observed in Ref.\cite{Rosa}. It stays at $\omega _{max}\sim
-t<0$ even when $\vec{k}$ crosses the Sh. F. S..

\subsubsection{Non-dispersive $T\rightarrow 0$ shadow feature in type
II configuration:}

The above considerations lead to the conclusion that only classical thermal
spin fluctuations that destroy long range order at $T\neq 0$ in two
dimensions can produce the above dispersive shadow peaks in both type I and
type II experiments. Quantum spin fluctuations $T\ll \omega _{sf}$ alone
cannot produce shadow effect in $A(\omega ,\vec{k}_{Fixed})$ measured in
type I ARPES, no matter how large is $\xi /a$. However, in the type II
configuration, ($\vec{k}${\em -dependence} of $A(\omega _{Fixed},\vec{k}))$,
quantum spin fluctuations can lead to a non-dispersive shadow feature at $T=0
$ when $\xi /a\gg 1$ (Ref.\cite{Chubukov}). The effect is due to the
enhancement of $\Sigma ^{\prime \prime }(\omega _{Fixed},\vec{k})$ on the
Sh.F.S. at finite $\omega $. For small $\omega $ the maximum in $\Sigma
^{\prime \prime }(\omega _{Fixed},\vec{k})$ translates into a maximum in $%
A(\omega _{Fixed},\vec{k})$ at $\vec{k}=k_{Sh.F.S.}$, since $A(\omega
_{Fixed},\vec{k})\propto \Sigma ^{\prime \prime }(\omega _{Fixed},\vec{k})$
for $\vec{k}$ far from the F.S.. Since $A(0,\vec{k}\neq \vec{k}_{F.S.})=0$
at $T=0$, one needs to increase $\omega $ to have a noticeable weight
for this maximum in the incoherent background. However, when $\omega $
increases the width of the maximum in $k$-space rapidly increases and above
a certain energy $\omega _0<t$, mainly determined by $(\xi /a)^2\omega
_{sf}$, the maximum disappears. This occurs because the effect from $\Sigma
^{\prime \prime }$ is offset by the contribution from the denominator in $%
A(\omega ,\vec{k})\approx \Sigma ^{\prime \prime }(\omega ,\vec{k})/{(\omega
-\tilde{\epsilon}(\vec{k}))}^2$. For $|\omega |>\omega _0$, the closer $\vec{%
k}$ is to the real F. S. the larger $A(\omega ,\vec{k})$ is. This is
illustrated in the inset of Fig.\ref{Diag1}. I found that the effect is
always very small even for large values of $\xi $ and $\xi ^2\omega _{sf}$.
For values of $\xi =2.3a$ and $\omega _{sf}\approx 10meV$ which are typical
for cuprates closed to the optimal doping \cite{MMP} it does not exist even
at very small $\omega \ll t$.

\subsection{Analytical results}

\subsubsection{Difference between classical and quantum regimes}

I now explain analytically why the effect of classical and quantum AFM
fluctuation on $A(\omega ,\vec{k})$ are qualitatively different. When $%
\omega _{sf}\ll T$ the effect of spin fluctuations on single-particle
properties can be considered quasistatic. Indeed, neglecting $\omega
^{\prime }$ in the energy denominator of the self-energy formula Eq. (\ref
{SigReal}) one obtains \cite{Lee}:

\begin{equation}  \label{SigS}
\Sigma (\omega ,\vec k)=g\bar{g}\int \frac{d^2q}{(2\pi )^2}\frac{S_{sp}(\vec %
q)}{\omega -\tilde \varepsilon (\vec k+\vec q)+i0}
\end{equation}
where $S_{sp}(\vec q)$ is the the static structure factor $S_{sp}(\vec q%
)=\int \chi _{sp}^{\prime \prime }(\omega ^{\prime },\vec q)\coth (\omega
^{\prime }/2T)d\omega ^{\prime }/2\pi $. The key point is that in the RC
regime, $S_{sp}(\vec q)$ has the {\em same} singular behavior as the static
susceptibility $\chi _{sp}^{\prime }(0,\vec q)$ because for $q\approx Q$, $%
S_{sp}(\vec q)\approx T\int (d\omega /\pi )\chi^{\prime \prime} (\omega ,%
\vec q)/\omega =T\chi _{sp}^{\prime }(0,\vec q)$. Since close to the AFM
instability $\chi _{sp}^{\prime }(0,\vec q)$ is a Lorentzian peaked at $q=Q$
, the static structure factor $S_{sp}(\vec q)$ has a Lorentzian maximum in
the regime $\omega _{sf}\ll T$ :

\begin{equation}  \label{Scla}
S_{sp}(\vec q)\propto \frac T{\xi ^{-2}+(\vec q-\vec{Q})^2}
\end{equation}
The presence of a sharp precursor of the Bragg peak in the RC regime leads
to the shadow peak in $A(\omega ,\vec k_{Fixed})$. The situation is
qualitatively different in the case $T\ll \omega _{sf}$ and $\xi /a\gg 1$.
In that case, $S_{sp}(\vec q)\approx \int_{0}^{\infty}
\chi^{\prime\prime}(\omega ,\vec q)d\omega /\pi $ and using Eq. (\ref{Susc1}%
) I obtain:

\begin{equation}  \label{Squ}
S_{sp}(\vec q)|_{T=0}\propto \ln [\xi ^{-2}+(\vec q-\vec{Q})^2]
\end{equation}
The logarithmic singularity of $S_{sp}(\vec q)$ in the case $T\ll \omega
_{sf}$ is integrable and does not lead to any divergence in $\Sigma (\omega ,%
\vec k)$ when $\xi \rightarrow \infty $. Consequently, there is no shadow
effects in $A(\omega ,\vec k_{Fixed})$, when electrons approach the AFM
instability $\xi \gg a$ at zero temperature $T=0$. This conclusion is
different from the result of Kampf and Schrieffer\cite{KShr} who used the
zero temperature formalism and found shadow bands for $\xi \approx 20a$. The
difference is due to the fact that the phenomenological susceptibility used
in Ref.\cite{KShr} was separable in $\vec q$ and $\omega $, $\chi
_{sp,KSh}=f(\vec q)g(\omega )$. In that case, the $\vec q-$dependence of the
static structure factor $S_{sp}(\vec q)$ is {\em always} the same as that of
the susceptibility $\chi _{sp}^{\prime }(0,\vec q)$. Consequently, the
static structure factor $S_{sp,KSh}(\vec q)$ has a Lorentzian-like peak even
when $T\ll \omega _{sf}$. The separability in $\omega $ and $\vec q$ of the
susceptibility is an oversimplication that can be partially justified only
in the regime $\omega _{sf}\ll T$. Thus, the results of Ref.\cite{KShr}
should not be applied to the zero-temperature paramagnetic state.

\subsubsection{{\it Conditions on $\xi (T)$} }

The asymptotic form of $\Sigma (\omega ,\vec{k})$ in the RC regime has been
obtained in Ref. \cite{VT1} for the case $n=1$, $t^{\prime }=0.$ The
generalization to the present case is straightforward and for $\Sigma
^{\prime \prime }$ I get, for $\xi >\xi _{th}=v_{\vec{k}{\bf +}\vec{Q}}/T$: 
\begin{equation}
\Sigma ^{\prime \prime }(\omega ,\vec{k})=-\frac{g^{\prime }T}{4\sqrt{\left[
\omega -\tilde{\varepsilon}(\vec{k}+\vec{Q})\right] ^2+v_{\vec{k}+\vec{Q}%
}^2\xi ^{-2}}}  \label{SigI}
\end{equation}
We see that $\Sigma ^{\prime \prime }(\omega ,\vec{k})$ has a maximum when $%
\omega =\tilde{\varepsilon}(\vec{k}+\vec{Q})$ with $\Sigma ^{\prime \prime }(%
\tilde{\varepsilon}(\vec{k}+\vec{Q}),\vec{k})=-g^{\prime }T\xi /4v_{\vec{k}+%
\vec{Q}}$. The effect is largest at the Van Hove point $v_k=0$, in which
case $\Sigma ^{\prime \prime }(\tilde{\varepsilon}(\vec{k}+\vec{Q}),\vec{k}%
)\propto \xi ^2$. Since away from $k=k_{F.S.}$, we have $A(\omega ,\vec{k}%
)\propto \Sigma ^{\prime \prime }(\omega ,\vec{k})$, we obtain for $\xi >\xi
_{th}$, the shadow feature discussed above. On the other hand, at the
crossing points $\tilde{\varepsilon}(\vec{k}_{cr}+\vec{Q})=\tilde{\varepsilon%
}(\vec{k}_{cr})=0$, $A(0,\vec{k})\propto 1/\Sigma ^{\prime \prime }(0,\vec{k}%
)$ and the enhancement in $\Sigma ^{\prime \prime }$ leads to the AFM
pseudogap with no quasiparticle peak inside. This is similar to what was
found in the case of perfect nesting ($n=1$, $t^{\prime }=0$) \cite{VT1},
except that in the latter case the pseudogap opens up over all the F.S.,
since the real F.S. and the Sh.F.S coincide (all points on the F.S. are
crossing points).

The real part of the self-energy can be obtained from Eq.(\ref{SigI}) using
Kramers-Kronig relation and has the form: 
\begin{equation}
\begin{array}{c}
\label{SigR}\Sigma ^{\prime }(\omega ,\vec k)=\frac{g^{\prime}T}{4\pi \sqrt{%
\left[ \omega -\tilde \varepsilon (\vec k+\vec Q)\right] ^2+v_{\vec k+\vec Q%
}^2\xi ^{-2}}} \\ 
\times \ln \left| \frac{\omega -\tilde \varepsilon (\vec k+\vec Q)+\sqrt{%
\left[ \omega -\tilde \varepsilon (\vec k+\vec Q)\right] ^2+v_{\vec k+\vec Q%
}^2\xi ^{-2}}}{\omega -\tilde \varepsilon (\vec k+\vec Q)-\sqrt{\left[
\omega -\tilde \varepsilon (\vec k+\vec Q)\right] ^2+v_{\vec k+\vec Q}^2\xi
^{-2}}}\right|
\end{array}
\end{equation}

To see how the shadow band solution appears, let us look at $\Sigma ^{\prime
}(\omega ,\vec k)$ at frequencies $|\omega -\tilde \varepsilon (\vec k{\bf +}%
\vec{Q})|\gg v_{\vec k{\bf +}\vec{Q}}/\xi $. In this case Eq. (\ref{SigR})
reduces to:

\begin{equation}
\Sigma ^{\prime }(\omega ,\vec{k})\approx \frac{g^{\prime }}{2\pi }\frac{%
T\ln \xi }{\omega -\tilde{\varepsilon}(\vec{k}+\vec{Q})}  \label{SigRT}
\end{equation}
When $T\ln \xi =const$ as $T\rightarrow 0$, the large value of $|\Sigma
^{\prime }|$ for $|\omega -\tilde{\varepsilon}(\vec{k}+\vec{Q})|\sim v_{\vec{%
k}{\bf +}\vec{Q}}/\xi $ leads to the appearance of two solutions for the
quasiparticle equation. The effect exists in $D=2$ in the R.~C. regime 
$\xi \propto \exp (const/T)$. Since in the weak-to-intermediate coupling
regime $T_X$ is of the order of the gap $\Delta $ in the ordered state, the
condition $\xi \gg \xi _{th}\equiv v_F/T$ can be also written as $\xi
\gg \xi _{coh}$, where $\xi _{coh}=\Delta /v_F$ is the coherence length in
the ordered state. In three dimensions precursors of AFM are very weak \cite
{VT1} because the integration of Eq.(\ref{SigS}) gives in 3D $\Sigma
^{\prime \prime }(\omega ,\vec{k}))\propto g^{\prime }T\ln \left[ [\omega -%
\tilde{\varepsilon}(\vec{k}+\vec{Q})]^2+v_{\vec{k}+\vec{Q}}^2\xi
^{-2}\right] $, which is much less singular then in two-dimensional case Eq.(%
\ref{SigI}). In the dimensions $D>3$ the precursors of AFM bands do not
appear even at the very point of the phase transition, because of the
large phase space in Eqs.(\ref{SigReal}),(\ref{SigS}). This result could be
expected on general physical grounds, since in three and larger dimensions
the gap is equal to zero at the N\'{e}el temperature. In contrast, in
two-dimensions classical thermal fluctuations suppress long-range order at
finite temperature and at the $T=0$ phase transition the system goes
directly into the ordered state with a {\em finite} gap. Finally, I
note that the one-dimensional case is, as usual, very special. For the most
recent studies of shadow bands in the 1D case see \cite{McKenzie,Penc}. 

\section{Conclusions}

I have shown that in 2D there are a few different antiferromagnetic
precursor effects in the spectral function $A(\omega ,\vec{k})$, which are
realized under different physical conditions. The conditions for dispersive
shadow features and shadow bands are given by $\omega _{sf}\ll T$ and $\xi
\gg \xi _{th}\equiv v_F/T$. The first conditions insures that the
short-range AFM order is {\em quasi-static}, which means that not only the
susceptibility $\chi _{sp}^{\prime }(0,\vec{q})$ has a Lorentzian form at $%
q\approx Q$ , but also the static structure factor $S_{sp}(\vec{q})$ has
this form. The second condition $\xi \gg \xi _{th}\equiv v_F/T$ implies that
the AFM correlation length should be compared with the thermal wave length $%
\xi _{th}=v_F/T$ of electrons, rather then with the
lattice constant $a$, as was assumed in some previous works on shadow bands.
The above conditions for dispersive shadow bands can be realized in
two-dimensions when at $T=0$ there is long-range AFM order. The situation is
favorable in two dimension because this is  the lower critical
dimension for the AFM instability and the mean field phase transition
is replaced by the crossover $T_X$ to the regime with very rapidly growing
correlation length $\xi \propto \exp (const/T)\gg v_F/T$ for $%
T\rightarrow 0$.

The dispersive shadow features and shadow bands are closely related to one
another. When the correlation length starts to increase faster than $%
\xi _{th}=v/T$ ($1/T^{1/2}$ at Van Hove point) the shadow bands appear
first at the hot spots, while shadow features appear at the wave vectors $%
\vec{k}$ away from the Fermi surface. When the correlation length
increases further the shadow feature is eventually replaced by the
shadow band. In this context, the thermally excited shadow feature could be
considered as a precursor of the shadow band.

The theory predicts no dispersive shadow features or shadow bands at
zero temperature no matter how large is the correlation length. I note that
the reason for the absence of dispersive shadow peaks at $T=0$ is similar to
the reason for the absence of shadow peaks in higher dimensions, namely,
large phase space. Indeed, in the Matsubara formalism the imaginary axis
plays the role of an additional dimension at $T=0$, since $%
T\sum_{\omega _n}(...)\rightarrow \int (...)d(i\omega )/2\pi $.

In addition to the dispersive shadow peaks which exist in both $\vec{k}$ and 
$\omega $ scans of the spectral function, there is also a non-dispersive
shadow feature \cite{Chubukov} in the $k-$scan only of $A(\omega
_{Fixed},\vec{k})$. The latter effect is due to the {\em dynamical} quantum
spin fluctuation and it exists at zero temperature and finite
frequency if $\xi $ is sufficiently large.

{\it Experimental implications:} Close to optimal doping the increase of $%
\xi $ with temperature is well described by the mean-field Curie-Weiss law $%
\xi \propto (A+BT)^{-1/2}$, i. e. slower than $1/T^{1/2}$ %
\onlinecite{MMP}. Under those conditions, the present theory does not
predict any dispersive shadow feature or shadow band. I also
found that the non-dispersive shadow feature due to quantum fluctuations\cite
{Chubukov} in $A(\omega _{Fixed},\vec{k})$ is always a very small effect. I
did not find such shadow feature for the conditions of the type
II experiment of Aebi et al.\cite{Aebi}. I conclude that, based on the
present approach, the ``shadow bands'' observed in B2212 close to optimal
doping \cite{Aebi,Rosa} cannot be explained as precursors of AFM bands. The
structural explanation of the ``shadow bands'' in optimally doped materials,
have been suggested in Ref.\cite{Chakravarty,Singh} (see also comment \cite
{Note3}).

I suggest that precursors of AFM bands should be searched for in strongly
underdoped materials in the RC regime $\xi \propto \exp (const/T)$ . These
bands should develop first at the crossing points $\vec k=\vec k_{cr}$ and
only much later along the diagonal direction $(0,0)-(\pi,\pi)$.

I am grateful to M. R. Norman and A. M.-S. Tremblay for many
suggestions  and stimulating discussions. I also acknowledge  useful
discussions with A. V. Chubukov and J. Schmalian. This work was
supported by the National Science Foundation (Grant
No. NSF-DMR-91-20000) through the Science and Technology Center for
Superconductivity.

\begin{figure}[tbp]
\caption{ The $\omega$-dependence of $A(\omega,\vec{k})$ for $\vec{k}={k}%
_{cr}=(0.825,0.175)\pi$. The inset show the F. S. (filled line) and the
shadow F. S. (dashed line).}
\label{cross}
\end{figure}
\begin{figure}[tbp]
\caption{ The $\omega$-dependence of $f(\omega)A(\omega,\vec{k})$ (upper
panel) and $\Sigma^{\prime\prime}(\omega,\vec{k})$ (lower panel) for $\vec{k}%
=(0.66,0.66)\pi$, that is slightly inside Sh. F. S. The inset shows the $%
\vec{k}$-dependence of $A(\omega,\vec{k})$ in the vicinity of $%
k_{Sh.F.S}=(0.63,0.63)\pi$ for $\omega=-0.03t$ ($\nabla$) and $%
\omega=-0.125t $ ($\Box$).}
\label{Diag1}
\end{figure}
\end{multicols}

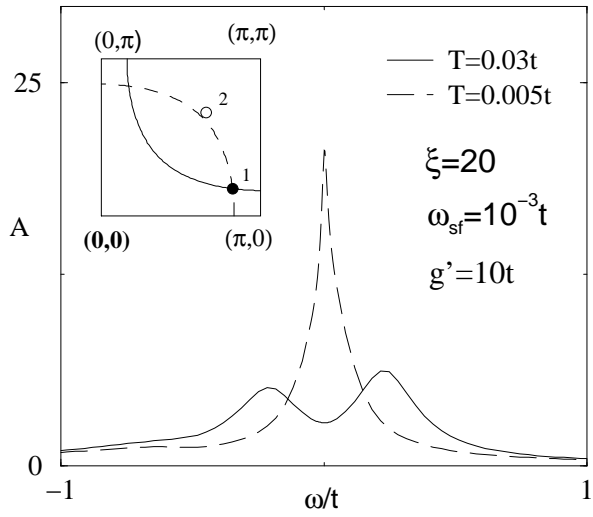
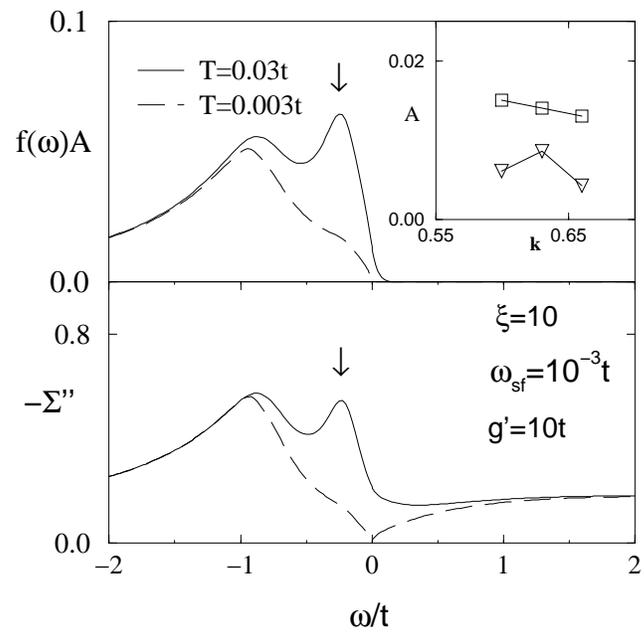
\end{document}